%% file: predictingconsent.tex
\documentclass[Afour,sageh,times]{sagej}

\usepackage{moreverb}
\usepackage[hyphens]{url}

\usepackage[colorlinks,bookmarksopen,bookmarksnumbered,citecolor=red,urlcolor=red]{hyperref}

\usepackage{tabularx}

\usepackage{rotating} 
\usepackage{multirow} 
\usepackage{dblfloatfix}

\newcommand\BibTeX{{\rmfamily B\kern-.05em \textsc{i\kern-.025em b}\kern-.08em

T\kern-.1667em\lower.7ex\hbox{E}\kern-.125emX}}

\def\volumeyear{2019}

\begin{document}

\runninghead{Norval and Henderson}

\title{Automating dynamic consent decisions for the processing of social media data in health research}

\author{Chris Norval\affilnum{1} and Tristan Henderson\affilnum{2}}

\affiliation{\affilnum{1}University of Cambridge, UK\\
	\affilnum{2}University of St Andrews, UK
}

\corrauth{Chris Norval, 
	Department of Computer Science and Technology,
	University of Cambridge,
	Cambridge,
	CB3~0FD, UK.
}

\email{chris.norval@cl.cam.ac.uk}

\begin{abstract}

Social media have become a rich source of data, particularly in health research. 
Yet, the use of such data raises significant ethical questions about the need for the informed consent of those being studied. 
Consent mechanisms, if even obtained, are typically broad and inflexible, or place a significant burden on the participant. 
Machine learning algorithms show much promise for facilitating a `middle ground’ approach: using trained models to predict and automate granular consent decisions. 
Such techniques, however, raise a myriad of follow-on ethical and technical considerations. 
In this paper, we present an exploratory user study ($n$ = 67) in which we find that we can predict the appropriate flow of health-related social media data with reasonable accuracy, while minimising undesired data leaks. 
We then attempt to deconstruct the findings of this study, identifying and discussing a number of real-world implications if such a technique were put into practice.
\end{abstract}


\maketitle

\input{Figures}

\input{Tables}

\section{Introduction}
Social media platforms, such as Facebook and Twitter, have become a significant and ubiquitous part of modern society. 
Approximately 77\% of UK-based online adults have a social media profile~\citep{ofcom:2018}. 
This data can be an invaluable resource for researchers with an interest in humans and society; 
Within a health context alone, research has suggested that such social data can act as a predictor for depression~\citep{moreno:2011,reece:2016}, suicide risk factors~\citep{choudhury:2016,jashinsky:2014}, mood changes~\citep{lee:2016}, flu outbreaks~\citep{li:2013} and problem drinking in US college students~\citep{moreno:2012}. 
Such findings can help researchers, practitioners and policy-makers gain insights into modern society, and facilitate the support of those in need who might otherwise have gone unnoticed.

Yet despite the potential advantages, those from whom the social data originated (who are, in effect, participants of such research) are not always informed about their data being used. 
Consequentially, they are not given the ability to decline or withdraw participation should they wish to -- cornerstones of conducting ethical research. 
Further, as these social repositories continue to grow, so too do (i) the attractiveness of mining and studying these data for research, and (ii) the associated challenges of gaining \emph{meaningful} and ongoing consent from a large group of people.

The topic of whether participant consent is needed for researching social media data (which can often be easy to access, or even publicly available) has generated opinions on both sides of the debate.
\cite{vitak:2016} documented survey responses from researchers, finding that many believed social data posted online are fair game to collect and analyse as long as they are accessible without signing in. 
Similar opinions were also found by~\cite{samuel2019}, when interviewing researchers about using social media data.
However, a number of other researchers have argued that just because personal information is made available online does not absolve one from their ethical responsibilities regarding participant consent~\citep{boyd:2012, chiauzzi2019, conway:2016, hunter2018, hutton:2015, rothstein2015, zimmer:2010}. 
For example, \cite{conway:2016} argue that the potential challenge to privacy occurs ``not in the reading or accessing of individual materials (publicly available as they are), but rather in the processing and dissemination of those materials in a way unintended''. 

Even putting aside questions over ethics, collecting and analysing social media data for research purposes without notifying the participant could have legal implications.
For example, the General Data Protection Regulation~\citep{gdpr2016} (GDPR), which came into effect on 25\textsuperscript{th} May 2018, provides increased data protection rights over citizens and residents of the European Union (EU).
Such rights include the `Right to be Informed' over the processing of personal data.\endnote{GDPR, Arts 13 and 14.}\textsuperscript{,}\endnote{Personal data is defined as ``any information relating to an identified or identifiable natural person (`data subject'); an identifiable natural person is one who can be identified, directly or indirectly, in particular by reference to an identifier such as a name, an identification number, location data, an online identifier or to one or more factors specific to the physical, physiological, genetic, mental, economic, cultural or social identity of that natural person;" GDPR, Art 4(1).}\textsuperscript{,}\endnote{Note that pseudonymised personal data is still considered personal data, due to the potential for future re-identification. GDPR, Recital 26.}
This places a legal responsibility on the `data controller'\endnote{The entities responsible for determining the purposes and means of processing the personal data. GDPR, Art 4(7).} to provide information (including the purposes of the processing, among other things) to any EU citizen or resident whose personal data is being processed.\endnote{While Article 14 does outline an exception where ``provision of such information proves impossible or would involve a disproportionate effort, in particular for processing for archiving purposes in the public interest, scientific or historical research purposes or statistical purposes", GDPR, Art 14(5)(b), the onus will be on the data controller to demonstrate the disproportionality with the supervisory authority, should a complaint be made.}
The GDPR's impact on research is yet to be fully realised, and is the subject of ongoing discussion within the academic community~\citep{chassang2017,mostert2016,mourby:governance,rumbold2017,veale2018}.

As such, obtaining the informed consent of research participants is an important ethical (and potentially legal) milestone to strive for. Yet, there is \emph{further} ongoing debate about whether consent should be one-off and broad, or continuous and specific~\citep{kaye:2014,luger:2013,morrison:2014,steinsbekk:2013}.
With the former, the participant lacks granular control and autonomy over what data is appropriate for sharing with the researcher, and for what purposes. 
The latter, in contrast, grants more control to the participant -- though the cost of this control is an increased burden placed on the participant to manage their data based on what they find appropriate. 
In short, a trade-off exists between configurability and convenience of consent mechanisms. 
\cite{baarslag:2017} argue that ``there is a pressing need for automating privacy negotiation that can make meaningful decisions on the user's behalf while minimizing their burden''.

In this paper, we explore machine learning as a mechanism for automating dynamic consent decisions -- striving for high predictive accuracy, while minimising the burden associated with repeated requests. 
We evaluate a number of algorithms and configurations, and outline potential implications of such an approach. 
Our work aims to be a case study into how the problem could be technically approached---warts and all---before then deconstructing and critiquing the approach taken in order to provide lessons and insights into this subject.

As such, our findings should be interpreted carefully, and with full hindsight of the intrinsic challenges of predicting consent as outlined in the Discussion section.
In this regard, the limitations and biases attempt to serve a cautionary tale for those with an interest in this topic.
We believe this work is of particular relevance to the research community, especially those with an interest in research ethics, user studies, social media, privacy, and fair and accountable machine learning. 

In this paper, we firstly outline and present the results of a user study ($n$ = 67) to predict dynamic consent decisions with regard to social media data in health research. 
We focus on this context given the potentially sensitive nature of medical data and its prevalence as an outcome variable in research involving social media data.
Secondly, we outline several considerations for model optimisation in a consent prediction context. 
And finally, we discuss some of the ethical, technical, and practical implications of using such a technique to predict consent decisions based on observations from our study. 
Our intention is not only to explore whether we \emph{can} predict participant consent decisions, but also under what circumstances we \emph{should} do so.

\section{Background} 
Informed consent is outlined by the \cite{apa:2014} as ``the process by which researchers working with human participants describe their research project and obtain the subjects' consent to participate in the research based on the subjects' understanding of the project's methods and goals''. 
It is ``widely seen as fundamental to medical and research ethics''~\citep{manson:2007}, and is a way in which the researcher can fulfil ethical responsibilities with regards to the protection of data, privacy, and the autonomy of the participant~\citep{morrison:2014}. 

Recent debates have questioned the suitability and practicality of traditional approaches to participant consent in the case of studying online communications. 
A common argument is that gaining consent from each participant studied in such an environment (which may include many thousands of unique accounts) may be impractical, or even impossible~\citep{hudson:2004,solberg:2010, willis:2017}, yet research has suggested that individuals in online environments generally do not approve of being studied without their consent~\citep{fiesler2018, hudson:2004}. 
Reviews of the literature into user attitudes toward the analysis of social media data for research have resulted in ``equivocal findings''~\citep{mikal:2016}. 
Despite this, the literature is increasingly reaching the consensus that consent should be sought in such cases~\citep{boyd:2012, chiauzzi2019, conway:2016, hunter2018, hutton:2015, rothstein2015, zimmer:2010}.
This is particularly the case when health-related data is involved~\citep{chiauzzi2019, conway:2016, fiesler2018, hunter2018, norval:2017, rothstein2015}.

\subsection{Broad, dynamic, and contextual consent}
Different approaches for obtaining consent have been defined. 
The de facto standard in research, referred to as `broad consent'~\citep{kaye:2014},\endnote{It is sometimes referred to in the literature simply as `informed consent', though this is likely more to do with it being the go-to default approach to `informed participant consent', rather than any statement that it leaves participants any more informed than other consent approaches we will discuss (such as dynamic consent).}
is typically a one-off, catch-all request conducted at the outset of a study. 
This approach has come under criticism in the literature for being unsuitable for research using online communications due to its lack of flexibility, transparency, and ongoing participant control~\citep{kaye:2014,luger:2013,morrison:2014}. 

As a result, many have argued for a more fluid approach, often referred to as `dynamic consent'~\citep{kaye:2014}. 
This approach involves giving the participant increased control over their data, including the ability to grant or revoke access to certain data for certain research~\citep{kaye:2014,luger:2013,steinsbekk:2013}. 
It promotes data re-use (with the knowledge and consent of the individual), and preferences can be modified over time on an ongoing basis~\citep{kaye:2014}. 
It has been described as superior with regard to autonomy, information, increased engagement, control, social robustness, and reciprocity~\citep{steinsbekk:2013}. 
Dynamic consent is not without its own criticisms, however. 
\cite{steinsbekk:2013} have argued that ``broad consent combined with competent ethics review and an active information strategy is a more sustainable solution''. 
One issue of dynamic consent is that repeated requests may lead to `consent fatigue', potentially risking attrition~\citep{hutton:2015,kaye:2014,morrison:2014,steinsbekk:2013}.

Some have subsequently sought a middle-ground approach. 
\cite{hutton:2015} have outlined `contextual integrity consent', based on \citeauthor{nissenbaum:2004}'s model of contextual integrity~\citep{nissenbaum:2004}. 
This approach looks to consider the contextual nature of appropriate data flow, taking into account factors such as the type of data, who is requesting it, and why it is being requested. 
\citeauthor{hutton:2015} also put forward a statistical approach to inferring when this decision might be automated, so as to not over-burden the participant.

\subsection{Automated consent procedures}
Predictive algorithms have been suggested as a potential solution to the problem of participant burden and informed consent in large-scale observational research studies~\citep{baarslag:2017, hutton:2015, jones2018,norval:2017}.
For example, \citeauthor{hutton:2015} investigated an approach involving the collection of consent decisions as the participants agreed or disagreed with sharing data for research purposes. 
As the participant continually answered these requests, the distribution of their answers for each data type was compared to the distribution of other participants. 
If the participant conformed with these ``norms'', the decision was automated, reducing participant burden. 
\citeauthor{hutton:2015}'s results suggest that such an approach may be an effective method of automating consent for those who are ``norm-conformant'', with such an approach defaulting to a `Dynamic Consent' approach when conformance could not be assumed.

\cite{gomer:2014} outline a proposal for a semi-autonomous agent using machine learning or a rule-based system for predicting consent decisions. 
Similar work has looked into automated negotiation agents for incentivised data sharing requests~\citep{aydogan:2017,baarslag:2017}, suggesting that such an approach can result in statistically higher accuracy compared to random chance~\citep{baarslag:2017}. 
Further calls for such an agent-based approach have been made with regard to consent in the age of the Internet of Things~\citep{schraefel:2017}.

While using statistical models to predict participant consent decisions may be possible, challenges with such an approach have been raised~\citep{jones2018,norval:2017}.
So far, much of the work in automated consent procedures have called for further research to be undertaken in order to understand the predictive capabilities of such approaches. 
In line with this, our work attempts to build on these findings by exploring the practical and technical implications of such an approach -- while simultaneously contributing to the debate on some of the wider ethical questions that such techniques could raise.

\section{User Study}
To explore how consent prediction might work in practice, we outline a scenario where we wish to predict whether an individual would find it appropriate for their social data (e.g. photos, status updates, page likes) to be shared with a given audience.
More specifically, we are interested in those who use Facebook (due to its relative popularity) for health purposes (be it for information retrieval, participating in online support groups, discussing medical conditions, etc.).
This offers an interesting and pertinent use-case, given the interest in social media for health research (outlined previously).
Our research hypothesis is that we can predict whether or not a given bit of social data should be shared (dependent variable) based on attributes of the social data itself and of its author (independent variable). 
This prediction might factor in, for example, the kind of social content in question (e.g. a photo, a status update, a page like), with whom it would be shared (i.e. the audience), and how willingly the individual has approved similar requests in the past.

To create and evaluate a predictive consent model for the above task, we designed a web-based study to collect a corpus of data. 
Participants were repeatedly asked whether they would find it appropriate to have different types of their social data shared with different hypothetical audiences within a medical context. 
From this dataset, we went on to train models to predict the appropriate flow of such data, and explored how the different factors of the model each influenced the consent decision.

\subsection{Application Development}
We developed a web application to conduct this data collection study remotely. 
This application made use of the Facebook API to present participants with social data from their Facebook profile, attempting to tap into the motivation to adequately protect their own social data~\citep{madejski:2012}.
However, this raised a number of considerations to overcome in order to conduct the study in an ethically aware way.

Given the potentially sensitive nature of the social data, we chose not to collect or analyse the social data itself. 
Rather, we collected contextual metadata about (i) the social data, (ii) the participant, and (iii) the request in question.
While collecting and analysing the data itself (e.g. through image recognition on pictures, sentiment analysis on status updates) may have led to improvements in classification accuracy over metadata alone, such an approach would have a separate set of ethical and practical implications.
As such, we consider such approaches to be outside the scope of this particular study.

To mitigate concerns over working with the Facebook data of participants, we used the PRISONER framework, described as an ``architecture for ethical and privacy-sensitive social network experiments''~\citep{hutton:2013}. 
This framework acted as middleware between our web application and the Facebook API, handling authentication along with the sanitation and collection of data during the study. 
PRISONER Configuration files specified which data should be temporarily accessible to the application, and which data should be stored for later retrieval by the researcher;
The former allowed our application to present the Facebook data in question to the participants, and the latter meant that only metadata was collected and accessible to the researchers.
Constraints of the Facebook API meant that only data created by the authenticated participant (i.e. no social data produced by their friends) was accessible to the application, and this did not include content posted to private groups.

We identified a number of potential contextual factors which could be of importance to the decision of whether data sharing was appropriate.
This included the participant (who was being asked), the data type (what kind of data was being requested), the audience (with whom it would be shared, and for what purpose).
Each item of social data could be categorised as one of the following Facebook data types, based on prior work~\citep{hutton:2015} in this area:\endnote{A sixth data type, identifying the name of a friend of the participant, was omitted due to restrictions of the Facebook API introduced since \citeauthor{hutton:2015}'s original study.}
(i) `Liked' Facebook pages, 
(ii) Status updates,
(iii) Location check-ins,
(iv) Photos,
(v) Photo albums.

We also outlined four hypothetical audiences within an online medical context to explore how this would impact the decision to share or withhold the social data. These audiences included:
(i) Researchers,
(ii) Clinicians,
(iii) Medical support group members
(iv) Members of the general public.
Explanations and examples of each of these hypothetical audiences were outlined to participants in an information sheet prior to starting the study, in an attempt to mitigate the abstract nature of the request.
While we recognise that this study is a narrow representation of consent decisions, we argue that it offers construct validity within a particular scope which has precedence in the literature (i.e. the work by \citealt{hutton:2015}).

\subsection{Recruitment}
The recruitment of participants proved to be one of the major challenges of this study. 
We began by identifying 50 UK-based medical support groups on Facebook.\endnote{Note that despite the health-oriented focus, we make no assumptions about any medical conditions that any participants may have.}   
We contacted the administrators of each of these groups, outlining the aims of our research, the nature of the study, measures taken to protect the privacy of participants, and finally asking if they would be willing to distribute information about the study to their members.
This approach was not particularly effective, and even resulted in a few negative responses -- further elaborated in the Discussion section of this paper. 
As such, we also created Facebook adverts targeting those with an interest in health conditions, distributed posters and leaflets at a digital health conference (to people who worked with patients), distributed an invitation to participate on Twitter, and used an academic recruitment website to promote our study. 
We specifically did not record the recruitment approach for any participant, and so cannot tie any specific respondent to a recruitment method or Facebook group. 
The study was open to all who met the recruitment criteria, and restrictions were thus not in place to control for demographic attributes. 
Further implications of our sample are critiqued in the Discussion section, and readers should consider this section carefully before extrapolating the implications of our analysis.

\subsection{Method}
In the first instance, our research proposal and study plan were scrutinised by the ethical review committee of the authors' institution, and approval to proceed with the research was granted.
Following approval, we recruited 100 UK-based adults over 18, who self-identified as using social media for health purposes, to participate in the research.
67 completed the study, with the remaining 33 participants either not finishing or actively withdrawing during the process. 
Participants who completed the study received a \pounds 10 amazon.co.uk gift card. 

The study consisted of three stages:

\subsubsection{Stage 1:} 
After reading the information page and granting consent, participants `signed in' to the experiment application using the Facebook API, granting the application access to their Facebook data.
Participants then completed a demographic questionnaire, collecting information about their age (bucketed, e.g. 18 -- 24, 25 -- 34), gender, level of education, Facebook privacy setting, Facebook friend count (rounded to the nearest 50 to obfuscate identity), and nationality. 
All of these values were optional. 
Participants were also asked for their email address, which was used both as a measure of uniqueness and for distributing the gift cards.

\subsubsection{Stage 2:} 
Participants were presented with a social item from their Facebook profile, and were asked if they would find it appropriate to have that social data shared with a randomised hypothetical audience (see Figure~\ref{fig:QuestionnaireOne}). 
This was repeated with a different combination of social data and audience, up to 100 times per participant.
The combination of data type and audience were pseudo-randomised to achieve a roughly even distribution of combinations ($\le$ 20 questions for each of the five data types). 
If the participant did not have enough social data for a given data type (e.g. they only had 10 check-ins), the rest of the questions for that data type were omitted. 
Data collected from each response included the hypothetical audience, the type of social data in question, the number of likes it received, the number of comments it received, the date and time published (rounded to the nearest hour to obfuscate identity), and the data's privacy setting.

\FigureQuestionnaireOne

\subsubsection{Stage 3:}
A response from Stage 2 was randomly selected and presented to the participant, and they were asked if that social data contained anything health-related. 
If the participant had specified that the data should \emph{not} have been shared, they would also be asked to specify their reasons for their answer via pre-defined check-boxes and an open text area (see Figure \ref{fig:QuestionnaireTwo}). 
The reasons given in the pre-defined check-boxes were partly adapted from work investigating reasons for self-censorship (offensive, uninteresting, self presentation)~\citep{das:2013,sleeper:2013}, and partly informed by contextual integrity (the content, the data type, the audience). 
This was repeated for up to 20 of the participant's prior answers.

\FigureQuestionnaireTwo

\section{Data Processing and Feature Selection}
This study resulted in a dataset of 4,660 consent decisions (stage 2)---of which, 1,027 had contextual reasoning (Stage 3)---from 67 participants. 
2,474 (53.1\%) of these consent decisions were seen as an acceptable flow of data (i.e. consent was granted). 
A link to our dataset and analysis code is included at the end of this paper.

First, participants were partitioned into three distinct groups: a training group ($n\textsubscript{pcpt}$ = 43), a testing group ($n\textsubscript{pcpt}$ = 12), and a validation group ($n\textsubscript{pcpt}$ = 12), controlling for the proportion of questions where consent was given per participant. 
The response data from these groups of participants made up 
a training set ($n\textsubscript{responses}$ = 3,031), 
a testing set ($n\textsubscript{responses}$ = 830), 
and a validation set ($n\textsubscript{responses}$ = 799). 
This ensured that each dataset consisted of the responses from a distinct group of participants, therefore allowing us to evaluate the generalisability of our approach when testing.
In each dataset, responses were `inner joined' with the demographic information of the participant who answered, so that each row of data contained all relevant information (dependent variables). 
Missing data for the optional demographic questions were re-assigned using the most frequent responses for each variable. 

Second, we computed aggregate data for each participant. 
We hypothesised that those who consented to a high proportion of data being shared with a particular audience would likely continue to share a similar proportion with that audience in the future. 
To evaluate this as a potential predictor, we randomly sampled 20\% of responses for each participant and used that data to calculate an overall estimated share proportion per participant, and an estimated share proportion for each audience type per participant. 
The data used to calculate these values was then discarded (since retaining this data for analysis would have raised issues over rigour). 
Finally, these computed proportions were `inner joined' with the remaining 80\% of response data per participant-audience combination such that each row of data contained (i) the share proportion for the participant in question and (ii) the share proportion for the participant-audience combination (in addition to all other dependent variables).

While these share proportion variables loosely attempted to proxy sharing behaviours over time, we recognise that our data collection was not longitudinal. 
Further research exploring how sharing behaviours change over long and short terms would therefore be highly complementary. 
The calculation of this potential predictor also comes at the cost of two downsides. 
First, we lose a fifth of the response data for all of our participants. 
And second, any model which makes use of these variables would need to calculate (or estimate) these share proportions as a prior value, known as the `Cold Start' problem~\citep{park:2009}.

\tbllogreg

\section{Results: Can we Predict Consent?}
These results are broken down into four stages. 
First, we present the output from a multilevel logistic regression analysis for the purposes of variable inference. 
Second, we train a number of established binary classification algorithms, evaluating the predictive performance of each. 
Next, we explore how some model optimisation techniques impact the performance of the models, discussing each within the context of automated participant consent. 
Finally, we select one model based on the above criteria and validate it with a set of previously unseen participants (the validation set), therefore evaluating how this model might generalise to new participants in the real world.

\textbf{Sampled sharing proportions appear to be a highly significant predictor of consent decisions.} 
We present the results of a multilevel logistic regression model, using backward stepwise elimination for variable selection,\endnote{This removed any variables that were found to be non-significant predictors of consent decisions in each step.} in Tables~\ref{table:results:logreg} and~\ref{table:results:logregtwo}. 
The first step contains only variables which relate to the context of the request.
The second step then includes demographic data of the participant. 
The third step then includes the sampled sharing proportions. 
These two variables are a highly significant predictor, increasing the effect size (McFadden's R\textsuperscript{2}) from .061 to .286.
We present all three steps for the purpose of inference, showing how the inclusion of different predictor variables influenced the model's efficacy.

\tbllogregtwo

Following the multilevel logistic regression results, we present a mathematical formula in Figure \ref{fig:fullformula} for predicting whether social media data should be shared. 
This formula includes all of the dependent variables which were found to be significant predictors in the logistic regression model, and is the set of dependent variables used when training subsequent models in this analysis.

\subsection{Considerations for Predictive Consent Models}
There are a number of factors that should be considered when predicting consent decisions. 
One such consideration is how different machine learning algorithms affect the predictive efficacy of the models on our testing dataset. 

To explore this, a selection of machine learning algorithms were trained and evaluated using the training and testing datasets respectively (using the formula outlined in Figure~\ref{fig:fullformula}). 
10-fold cross-validation was used to reduce the risk of over-fitting, and hyperparameters were optimised via grid search. 
All analysis was performed in R using the `caret' package. 
The results of these evaluations are presented in Table \ref{table:results:modellist}, and their ROC curves are illustrated in Figure \ref{fig:testroc}.

\FigureFullFormula

\tblmodellist

\textbf{A trade-off exists between false positives and false negatives, and these can be re-balanced.}
A further consideration to binary classification models is that the importance of different performance metrics may often depend heavily on the context of what is being predicted. 
For example, a medical professional making clinical diagnoses may weigh a false negative (a sick person incorrectly classified as healthy) to be significantly more `costly' than a false positive (a healthy person incorrectly classified as sick).
In line with this, we can therefore explore how our models might perform when taking optimisation steps to minimise the number of false positives (data is shared when it should have been withheld) at the expense of false negatives (data is withheld when it should have been shared).

One method of achieving this cost-sensitive classification involves adjusting the probability threshold for where one class should be selected over the other~\citep{kuhn:2013,sinha:2004,zhao:2008}.
Given that we wish to lower false positives, we can adjust the probability threshold of each of our models such that the calculated probability of a \emph{Share} outcome must be higher than this threshold value in order for \emph{Share} to be predicted. 
This could be increased from 50\% to 95\%, resulting in a more conservative model, with fewer instances where data is leaked undesirably at the cost of lower model sensitivity (i.e. correctly identifying fewer instances where the data should be shared). 

\FigureTestROC

The predictions generated during the cross-validation training process were used to retrieve the probability threshold value where specificity was approximately .95. 
This represents the point at which approximately 95\% of data that should not be shared would be correctly withheld. 
We then evaluated each of our models' ability to predict on the testing set when this threshold was used, illustrated in Figure~\ref{fig:tpvsfp}.
We found that using this threshold adjustment technique reduced the False Positive Rate to approximately 5\% for many of our models, though lowering the sensitivity in the process.

\textbf{Data Leaks can be eliminated at the cost of burdening the participant.} 
If false positives are seen as entirely unacceptable, the participant could simply be consulted every time a `Share' outcome is predicted -- essentially eliminating data leaks through erroneous share predictions.
Taking this approach would mean that the proportion of requests requiring participant input would be dependent on how frequently the model predicts that data should be shared.
More conservative models would result in lower burden, but also lower sensitivity (true positives). 
In the worst case (i.e. all data is predicted as `Share'), this approach would match the burden associated with non-automated dynamic consent (the participant is asked whether each item should be shared).

\textbf{A model with fewer predictors may have minimal performance impacts, while being more privacy-aware.}
We can investigate the performance impact associated with following `data minimisation'\endnote{GDPR, Art 5(1)(c).} principles---collecting and processing less data---by removing some of the predictors from our models.
These `Minimised' models may be more acceptable in situations where participants are concerned about their data being accessed or processed to make a consent prediction.
Given that the Share Proportion variables were highly significant predictors of consent decisions (Table~\ref{table:results:logregtwo}), we define a `minimised' formula comprising of the share proportions as predictors, as shown in Figure \ref{fig:minformula}.

\FigureMinFormula

\TPvsFP

Given its previous performance (Table~\ref{table:results:modellist}), we compare the impact of using this formula on the na\"{\i}ve Bayes models for both threshold-adjusted and non-adjusted approaches.
Results are presented in Table~\ref{table:results:optimisationlist}.
Given the reduction of information, one might expect that the minimised formula would perform notably worse than the `Full' formula, outlined in Figure \ref{fig:fullformula}. 
Results of this comparison, however, suggest that this did not appear to be the case. 
This raises follow-up questions about situations in which participants might prefer the minimised model, and what drop in predictive performance might be seen as acceptable.
The use of share proportions as the only predictors, however, does have technical limitations -- as will be outlined in the Discussion section.

\subsection{Selecting and Evaluating a Model}
So far, we have evaluated multiple machine learning models (various algorithms; with and without adjusting the probability threshold; two mathematical formulas). 
However, performing multiple evaluations on the test set increases the likelihood of finding good results by chance. 
Selecting one `best-performing' model and evaluating it with the (previously unseen) validation set will give us more confidence in the generalisability of our findings. 
Yet, what constitutes a `best-performing' model is heavily dependent on the task in question, and requires us to determine a context under which we would opt for certain attributes of the model (threshold adjusted, full or minimal formula, etc.).

To evaluate a single model with the validation dataset, we select one that fits the criteria based on what we believe is a promising use-case of this consent prediction technique: Automating the dynamic consent process in a research databank containing social media data. 
For this hypothetical use-case, participants can answer a few example questions to configure their sharing preferences (i.e. calculate their share proportion variables), and then opt-in to having their social data shared with researchers or clinicians automatically. 
The system determines which specific data should be accessible to different audiences as requests for research participants are made. 
We select a model which (i) Minimises data leaks to approximately 5\% of requests where consent should not be given, (ii) minimises participant burden beyond the initial calculation of sharing preferences, and (iii) uses the full set of predictor variables that the databank would already have access to.

Based on this criteria, we choose the threshold adjusted na\"{\i}ve Bayes model using the set of predictors outlined in Figure \ref{fig:fullformula} to evaluate our validation dataset of previously unseen participants. 
The confusion matrix of this evaluation is presented in Table~\ref{table:results:confusionmatrix}. 
Based on these results, 97.0\% of consent requests which should not have been granted were correctly refused.
This, however, comes at the cost of sensitivity -- only 32.7\% of requests where consent would have been granted were correctly predicted as such.
Overall accuracy was 65.3\%.

\tbloptimisationlist

\section{Discussion: Should we Predict  Consent?}\label{qual}
The question of whether or not algorithms \emph{should} predict participant consent decisions is, of course, complex. 
While an answer to this question is beyond the reach of a single paper, we do believe that it is a question that those with an interest in research ethics need to start thinking about -- particularly as researchers have already began to suggest such an approach~\citep{baarslag:2017,hutton:2015,jones2018,norval:2017}.
Automated consent prediction requires a substantial discussion within the research community, along with further research, in order to better understand some of the potential implications.
In this section, we attempt to further this topic by outlining some observations raised from our study, and discuss a few considerations which we believe are pertinent to this topic. 

\textbf{There are \emph{very} serious differences between consent prediction with \& without the participant's permission.}
We selected a model for validation based on a hypothetical databank use case.
In this scenario, the participant would opt-in (broad consent) to a system which automated their granular sharing decisions (dynamic consent) -- with their full oversight. 
This is a very different scenario to predicting whether someone would consent to participation without their prior knowledge, and without first obtaining their permission.
Predicting `overall' consent (i.e. deciding what social profiles to scrape) without that participant's permission would not absolve a researcher of the ethical and legal implications outlined in the Introduction section of this paper. 
A clear distinction should be made between instances where overall consent is, and is not, obtained.
As such, we argue that \emph{any form of consent prediction should not be used without the participant's prior knowledge and informed permission}.

\textbf{Prediction need not necessarily mean automation.}
Of course, also relevant is the purposes for which consent prediction is performed. 
There are differences between predicting consent decisions and then seeking confirmation from the participant (in the form of active and ongoing dynamic consent) as opposed to outright automating the consent procedure. 
Of course, regardless, the participants in question should be aware, consenting, and involved in this discussion prior to any predictions or automation taking place.

\textbf{Most support groups were not willing to share information about this study with their members.}
Participant recruitment was a significant challenge for our study.
Firstly, while the study was designed with careful consideration of the ethical and privacy-related implications of handling participants' social data, we received a low response rate from moderators of the UK-based medical support groups on Facebook. 
50 groups were contacted, 15 replied (30\%), 9 of which were willing to allow the study to be shared with their members. 
The 6 groups who responded negatively either expressed the desire to protect their members, specified that their group was not meant for research, or raised specific concerns over the study accessing the members' Facebook data.\endnote{It may be worth noting that this study was conducted before the Cambridge Analytica revelations involving allegations of Facebook data misuse through `personality test' Facebook apps designed to harvest data~\citep{cadwalladr2018}; the evidence is mixed as to how well such studies are addressed by research ethics committees~\citep{schneble:internet-mediated} and indeed the personality test studies did not have approval~\citep{weaver:rejected}. Our study did, however, take place after the earlier `emotional contagion' Facebook research controversy~\citep{flick:facebook}.}
As a result, the number of participants recruited for this study was less than we had originally intended.

\tblconfusionmatrix

\textbf{Requests for publicly available data were not always granted, confronting the argument of `implied consent'.}
We have previously discussed the argument that just because social media data might be publicly available does not mean that it is appropriate to scrape and analyse without consent \citep{boyd:2012, conway:2016, hutton:2015, zimmer:2010}.
Our data appears to corroborate this viewpoint. 
Of data which included privacy/visibility settings (only Albums and Notes, as per Facebook's API), 15.4\% was publicly avaiable. 
Of this publicly available data, 41.2\% did not receive participant consent deeming it appropriate for sharing. 
When looking only at instances where the question requested sharing with researchers, this value was 35.7\%. 
In other words, this Facebook data were publicly available, however, the participants did not find it appropriate for that data to be accessed and used by researchers in a healthcare context. 

\textbf{31\% of data not shared with researchers was, in part, due to being perceived as uninteresting.}
We can look at data from Stage 3 of the study to explore reasons why consent was not given, and particularly we can look at cases where data was not shared with researchers. 
Of data which was withheld, `It is uninteresting' was checked one third of the time. 
This could indicate a misconception by participants that data they deem to be irrelevant isn't valuable to researchers.
Among other major reasons, 33\% was due to the data being personal, 22\% because it would be shared with a researcher, 16\% due to what the social data contained, and 13\% due to the type of data in question. 

\textbf{Requiring consent can lead to selection bias, and this may be an unavoidable cost of ethical research.} 
There is inherent selection bias in any data collected with consent~\citep{fiesler2018}.
There is irony in that a recognised limitation of our study is selection bias -- a consequence which may not have been as much of an issue if less scrupulous methods for data collection were utilised.
As an example, the vast majority of our respondents were female (86.6\%), though it is not distinguishable if this is due to females being more likely to engage with medical content on social media, or more likely to participate in research, or any other reasons leading to the gender disparity.
As such, this raises questions over how well our model would generalise if deployed to different types of individuals.

By asking people if they are willing to participate in a data collection study, the subsequent models generated from that dataset are based solely around the types of people who are likely to agree to take part in such research -- those who were uncomfortable in participating are therefore not represented. 
Nevertheless, researchers need to be aware of the implications, develop alternative recruitment strategies which mitigate this bias, and argue the case that ethical approaches are still worth pursuing.

\textbf{Performance metrics are only part of the picture, and accurate models may lead to further selection bias.}
During the model comparison process, we tested a model which used the minimised formula with a threshold adjustment, outlined in Table \ref{table:results:optimisationlist}.
This model identified 57\% of data that should be shared while reducing data leaks to 6.9\% of data which should not. 
From a purely metric-oriented perspective, it appeared to perform respectively.
However, in line with the previous subsection, the generalisability of any such approach to consent prediction should be carefully considered and scrutinised.
Further investigation into our predictive model identified the potential for unintended consequences, which could have significant research implications, if it were put into practice in the real-world. 

Since the sole predictor of our model was the participant's share proportions, and the model was threshold adjusted, the decision of whether consent was granted ended up depending entirely on the participant having previously shared a very high proportion of requests (e.g. \textgreater\textasciitilde95\%).
A very small number of participants with a high share proportion had all of their data shared, whereas participants with a lower share proportion (\textless\textasciitilde95\%) had none of theirs shared.
This led to a model which appeared to have relatively good performance metrics (and few data leaks), however, any research which utilised a dataset generated from such a model would be at risk of severe selection bias. 
Further, this share proportion was calculated from a fixed point in time -- no longitudinal data was collected. 
As a result, it would be easy to assume that it was a proxy of sharing behaviours over time.
However, given that it is perfectly feasible for sharing behaviours to change over time, this could lead to incorrect predictions being made if deployed in the real world.

In short, these observations illustrate how conducting solely metric-driven approaches to predicting consent decisions could lead to widely different findings if deployed in the real world.
Our model performed seemingly well at a tertiary glance -- though upon further scrutiny, we have identified several points of consideration.
We raise these points to emphasise that performance metrics for consent prediction aren't everything, and further investigation is vital before deploying a model with such consequences.
We intend for these insights to act as a cautionary tale, outlining some of the considerations that those developing consent prediction systems must consider in order to ensure accurate and ethical deployments going forward.

\section{Best Practices}
Machine learning practitioners (whether researchers, data scientists, or hobbyists) are facing increasing calls for ethical, transparent, and accountable practices~\citep{singh:2019}. 
We have argued that this is particularly true for predicting when social media data should be used for health research, and so-called `implied consent' cannot be assumed. 
More widely, we argue that any form of consent prediction should not be used without the participant's prior knowledge and informed permission. 
Further, predicting granular consent decisions (i.e. where permission has been given for consent prediction to take place) raises a number of considerations, which we have outlined in the Discussion section. 

In all, researchers must consider the wider context in which these models are deployed. 
While performance metrics may tell a part of the story about the efficacy of a predictive model, it may not accurately reflect the challenges it will face when deployed in practice. 
Seemingly high-performing models may predict poorly under particular circumstances, or with certain cohorts, and this may not become apparent until data leaks occur and harms result. 
This paper intends to lay the foundations for discussing some of these considerations, although it is not an exhaustive list, and careful circumspection is therefore advised.

\section{Research Agenda}
The present work is outlined as one exploratory case study into consent prediction for social media data in health research. 
As such, there are many intriguing areas for follow-on research. 
Firstly, we are careful to stress that our study contains no longitudinal elements. 
Given our finding that the proportion of social media data that an individual shares is a highly significant predictor of granular consent decisions (Table~\ref{table:results:logregtwo}), whether and how this predictor changes over time is an outstanding research question. 
Indeed, we believe that this is of paramount importance to the concept of consent prediction -- consent, after all, is fluid. 
A greater understanding of the longitudinal implications of this predictor could therefore help prevent the risks of data leakage if such an approach were ever deployed.

Additionally, as discussed, our research specifically explored predictions based on contextual factors -- loosely based on Nissenbaum's model of contextual integrity~\citep{nissenbaum:2004}. 
Follow-up work which explores consent prediction using more involved means of data mining (e.g. through image recognition on pictures, sentiment analysis on status updates)---providing it can be done in an ethically appropriate way---may lead to significant increases in accuracy. 
This would have strong implications on the degree to which such consent mechanisms can be relied upon. 
Based on our findings, we have argued that our approach should be looked at tentatively. 
However, alternate approaches which boast exceptionally high predictive accuracy may offer a way forward for consent automation in certain situations.

Other platforms also offer further opportunities for research, given that health information online is not constrained to Facebook. 
Our study could easily be replicated on other social media platforms with health-related communities, such as Twitter and Reddit. 
Given the contextual nature of requests, this could go on to help provide researchers with a richer understanding of what participants deem appropriate, where differences exist, and what accuracy can be achieved across these platforms.

\section{Educational Implications}
Raising awareness of this topic, along with some of the pitfalls and challenges which may not have otherwise been apparent, will better ensure that those serving on ethical committees can better scrutinise large-scale social media research with automated consent mechanisms. 
We believe that this is good for researchers and participants alike.

Additionally, we found that one third of publicly available data was deemed by the participants to be inappropriate for sharing with researchers---an empirical confrontation to the belief that publicly available data are fair game---which corroborates the viewpoint that making data public does not equate to implied consent. 
This has serious implications for researchers, both established and in training, about what participants expect of them.

We also found that 31\% of share requests with the `Researcher' audience-type were refused because they were deemed to be `uninteresting' by the participant. 
This could indicate potentially lost opportunities for researchers to recruit participants who have misconceptions, reservations, and/or concerns about how their data are used. 
We believe that this highlights the importance of ensuring that (to the greatest extent possible) participants are made aware of how their data is used and processed by researchers. 
This may include providing illustrative examples of the types of data analysed by the researcher (e.g. demonstrating that participant data is processed in aggregate, as opposed to directly identifiable), or indicating what the researchers hope to gain from access to their data (e.g. looking at patterns in sharing behaviour, as opposed to content analysis).

\section{Conclusion}
Social media platforms have continued to grow in popularity, making them a valuable resource for researchers.
Yet, such social data is often scraped and analysed without the explicit consent of the participant(s) in question, raising significant ethical and legal implications about such research.
Even when consent is obtained, current mechanisms are either broad and inflexible, or put the burden of continual data management onto the participant.
We have \emph{tenatively} explored a mechanism which attempts to combine broad and dynamic consent approaches by using machine learning to predict appropriate data flow, and we present results of several predictive models, finding respectable accuracy.

Yet, consent prediction remains a highly contextual and complex task, with pertinent ethical and legal implications.
We highlight several considerations that those exploring consent prediction systems should consider.
For example, while possible to obtain reasonable performance metrics in studies, obtaining representative samples and evaluating performance over time are vital -- though pitfalls associated with these biases may not be overly obvious. 
More widely, we want to raise awareness of such considerations, so that consent prediction---if and when it ever becomes commonplace---might better represent the intentions of our participants going forward.

\begin{acks}

We would like to greatly thank the administrators of the groups who helped us by sharing our study with their members, and all participants who agreed to take part. 
This work was supported by the Wellcome Trust [UNS19427]. 
The first author has since received funding from Microsoft through the Microsoft Cloud Computing Research Centre (MCCRC).

\end{acks}

\section{Dataset}
Our dataset and analysis code are available at: \url{https://github.com/cnorval/automating-dynamic-consent-dataset}

\theendnotes

\bibliographystyle{SageH}

\bibliography{predictingconsent}

\end{document}

%% file: Figures.tex
%
%
%

\newcommand{\FigureQuestionnaireOne}{
\begin{figure}[t]
\centering
\setlength\fboxsep{0pt} 
\fbox{\includegraphics[width=.9\linewidth]{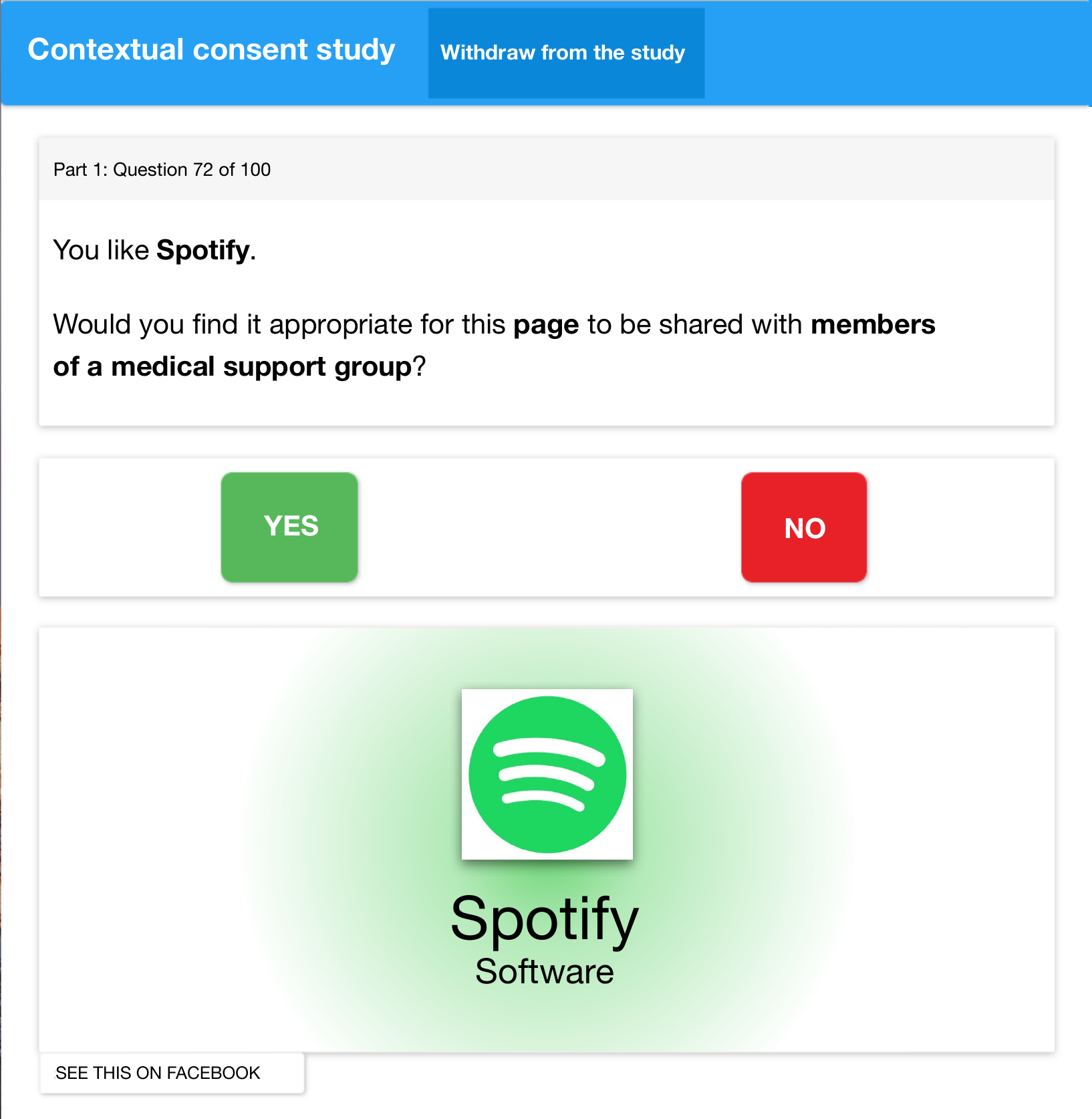}}
\caption{A question from Stage 2 of the study. Participants 
answer whether they would find it appropriate for a given social item from 
their Facebook profile to be shared with a hypothetical audience.}
\label{fig:QuestionnaireOne}
\end{figure}
}

%
%
%

\newcommand{\FigureQuestionnaireTwo}{
\begin{figure}[t]
\centering
\setlength\fboxsep{0pt} 
\fbox{\includegraphics[width=.9\linewidth]{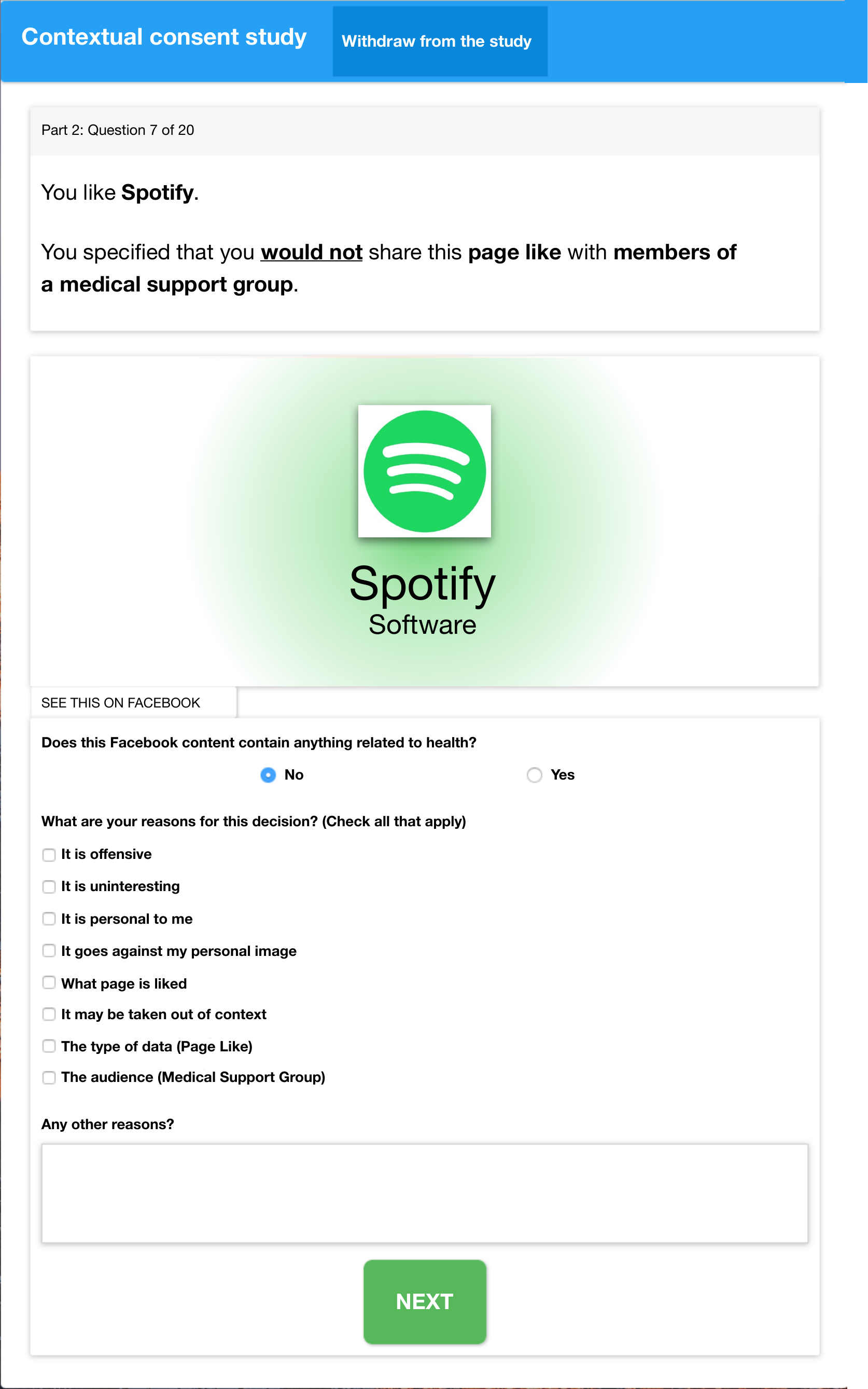}}
\caption{A question from Stage 3 of the study. Participants 
can specify the reasons why they chose not to share a social item with an audience.}
\label{fig:QuestionnaireTwo}
\end{figure}
}

\newcommand{\FigureFullFormula}{
\begin{figure}[!h]
	$ is\ shared \ \approx \ social\ data\ type\ +\ participant's\ education\ +\ number\ of\ friends\ +\ sampled\ overall\ share\ proportion\ +\ sampled\ share\ proportion\ with\ that\ audience $
	\caption{The formula showing the selection of predictor variables used in follow-up analysis.}
	\label{fig:fullformula}
\end{figure}
}

\newcommand{\FigureMinFormula}{
\begin{figure}[!h]
	$ is\ shared \ \approx \  proportion\ shared\ total\ +\ proportion\ shared\ with\ that\ audience $
	\caption{The `Minimised' formula.}
	\label{fig:minformula}
\end{figure}
}

%
%
%

\newcommand{\FigureTestROC}{
	\begin{figure}[!t]
		\centering
		\includegraphics[width=.8\linewidth]{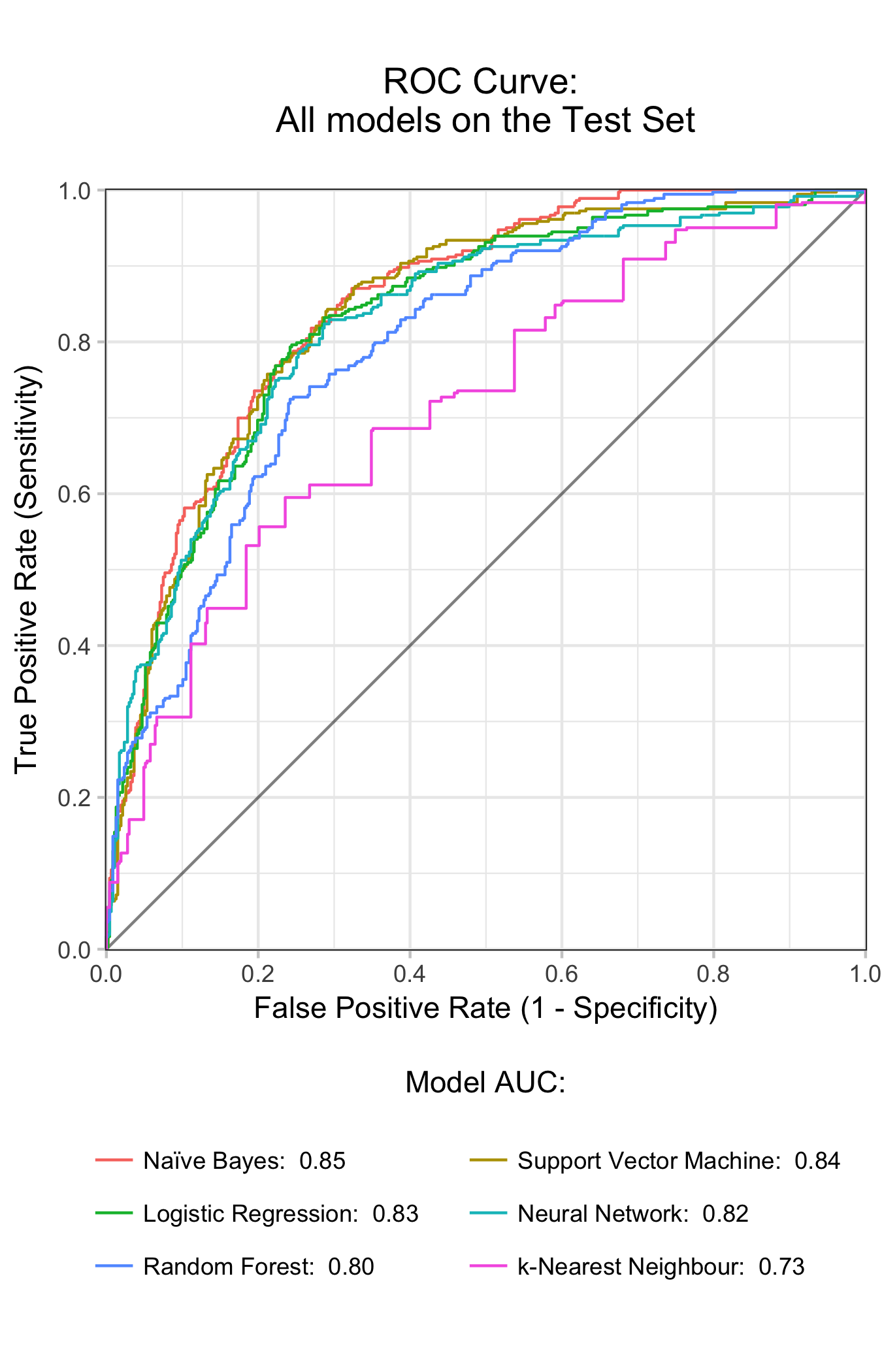}
		\caption{A collection of ROC curves of our models on the test set. Many 
		of these curves appear very similar, with little variance in shape or skew.}
		\label{fig:testroc}
	\end{figure}
}

%
%
%

\newcommand{\TPvsFP}{
	\begin{figure*}[th]
		\centering
		\includegraphics[width=.8\linewidth]{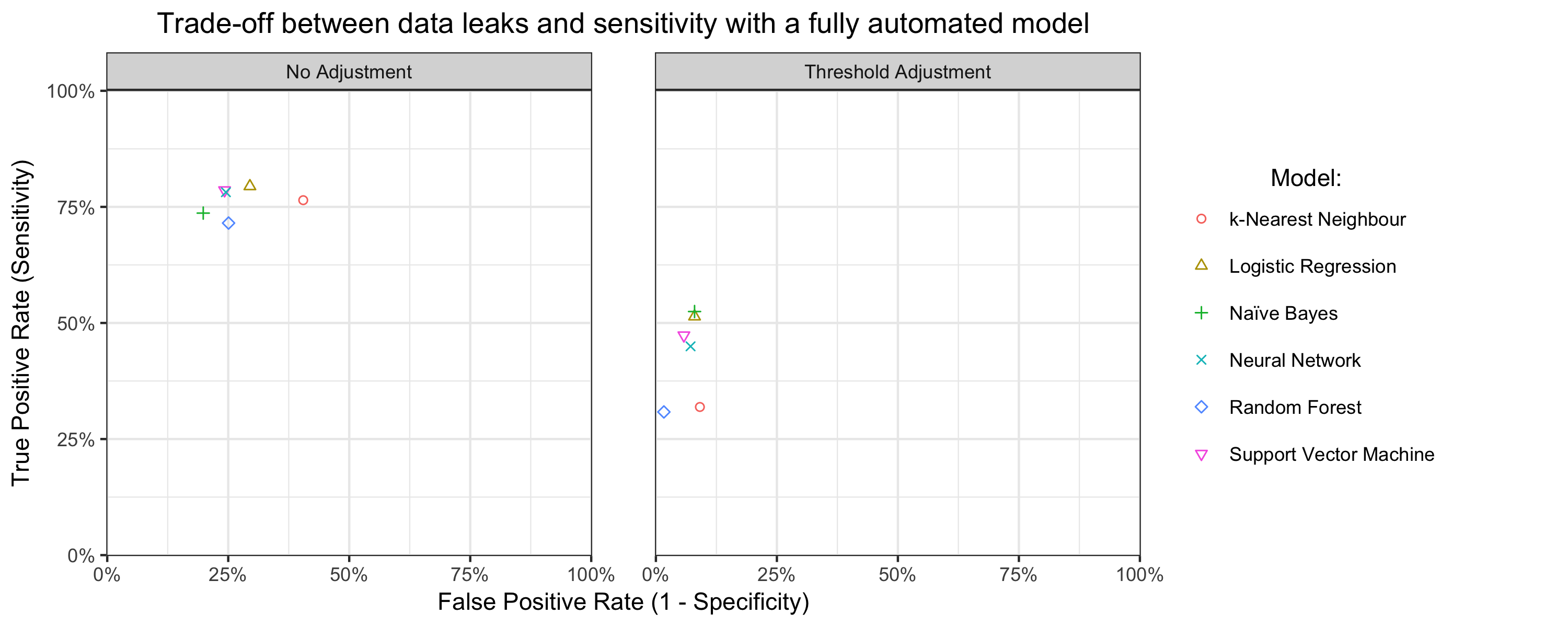}
		\caption{A comparison of the trade-offs between the sensitivity and the proportion of False Positives for different models, both with and without threshold adjustment.
		The threshold adjusted models attempt to reduce the instances of data leaks (false positives) at the cost of lower sensitivity --- which may be an acceptable sacrifice in some contexts.}
		\label{fig:tpvsfp}
	\end{figure*}
}

%% file: Tables.tex
\newcommand{\tbllogreg}{
	\renewcommand{\arraystretch}{1.0}
	\begin{table}[!b]
		\caption{The first two steps of the multilevel logistic regression model to predict consent decisions. With only contextual and demographic factors, Step 2 does not capture much variance (McFadden's R\textsuperscript{2} = .06).}
		\label{table:results:logreg} 	
		
		\centering
		\footnotesize
		
		\begin{tabularx}{\linewidth}{Xccl} 
			
			\toprule
			\textbf{Coefficient} & \textbf{Value} & \textbf{SE} & \hspace{.4cm}\textbf{p} 
			\hspace{.6cm}
			\\ 
			
			\midrule
			
		\end{tabularx}

		\vspace{.1cm}
		
		\flushleft{\textbf{Step 1:} Contextual}

		\begin{tabularx}{\linewidth}{Xccl} 
			\hspace{.5cm}Intercept			   
			& -0.64 & 0.13 & 0.000 *** \\ 
			
			\hspace{.5cm}Data type & ~ & ~ & ~ \\
			\hspace{1cm}Checkin					   
			& 0.47 & 0.13 & 0.000 *** \\ 
			\hspace{1cm}Like 						   
			& 1.11 & 0.12 & 0.000 *** \\ 
			\hspace{1cm}Note 						 
			& 0.67 & 0.12 & 0.000 *** \\ 
			\hspace{1cm}Photo 						
			& 0.40 & 0.12 & 0.001 *** \\
			
			\hspace{.5cm}Audience type & ~ & ~ & ~ \\
			\hspace{1cm}Group					   
			& 0.06 & 0.10 & 0.569  \\ 
			\hspace{1cm}Public 						   
			& 0.02 & 0.11 & 0.813  \\ 
			\hspace{1cm}Researcher 						 
			& 0.32 & 0.11 & 0.003 ** \\ 
			
			\hspace{.5cm}Published Time  & ~ & ~ & ~ \\
			\hspace{1cm}Evening					   
			& 0.04 & 0.11 & 0.728  \\ 
			\hspace{1cm}Morning				   
			& 0.19 & 0.10 & 0.069  \\ 
			\hspace{1cm}Night				   
			& 0.21 & 0.10 & 0.038 * \\

		\end{tabularx}
		
		\vspace{.1cm}
		\hfill McFadden's R\(^2\) = .027
		\vspace{.25cm}

		\flushleft{\textbf{Step 2:} Contextual \& Demographic}
		
		\begin{tabularx}{\linewidth}{Xccl}

			\hspace{.5cm}Intercept			        
			& -2.45 & 0.33 & 0.000 *** \\ 
			
			\hspace{.5cm}Data type & ~ & ~ & ~ \\
			\hspace{1cm}Checkin					   
			& 0.46 & 0.13 & 0.000 *** \\ 
			\hspace{1cm}Like 						   
			& 1.14 & 0.12 & 0.000 *** \\ 
			\hspace{1cm}Note 						 
			& 0.67 & 0.12 & 0.000 *** \\ 
			\hspace{1cm}Photo 						
			& 0.41 & 0.12 & 0.001 *** \\ 
			
			\hspace{.5cm}Audience type & ~ & ~ & ~ \\
			\hspace{1cm}Group					   
			& 0.04 & 0.11 & 0.689 \\ 
			\hspace{1cm}Public 						   
			& -0.01 & 0.11 & 0.907 \\ 
			\hspace{1cm}Researcher 						 
			& 0.33 & 0.11 & 0.003 ** \\ 
			
			\hspace{.5cm}Published Time  & ~ & ~ & ~ \\
			\hspace{1cm}Evening					   
			& 0.02 & 0.11 & 0.843  \\ 
			\hspace{1cm}Morning				   
			& 0.18 & 0.11 & 0.080  \\ 
			\hspace{1cm}Night				   
			& 0.22 & 0.10 & 0.037 * \\ 
			
			\hspace{.5cm}Education & ~ & ~ & ~ \\
			\hspace{1cm}High School  					   
			& 0.88 & 0.12 & 0.000 *** \\ 
			\hspace{1cm}Undergraduate Degree 						
			& 0.74 & 0.10 & 0.000 *** \\ 
			\hspace{1cm}Postgraduate Degree 			    
			& 0.72 & 0.13 & 0.000 *** \\ 
			
			\hspace{.5cm}Profile Visibility & ~ & ~ & ~ \\
			\hspace{1cm}Friends Only					   
			& 1.01 & 0.30 & 0.001 *** \\ 
			\hspace{1cm}Other					   
			& 0.26 & 0.39 & 0.514  \\ 
			\hspace{1cm}Public
			& 2.57 & 0.58 & 0.000 *** \\ 	
			
			\hspace{.5cm}Number of Friends
			& 0.00 & 0.00 & 0.000 *** \\

		\end{tabularx}
		
		\vspace{.1cm}
		\hfill McFadden's R\(^2\) = .061
		\vspace{.1cm}

		\hrulefill	
		
		\vspace{.1cm}
		
		\centering
		* p \textless~.05;\hspace{1.2em} 
		** p \textless~.01;\hspace{1.2em} 
		*** p  \textless~.001

	\end{table}
}

\newcommand{\tbllogregtwo}{
	\renewcommand{\arraystretch}{1.0}
	\begin{table}[!bh]
		\caption{The third step of the multilevel logistic regression model, adding in past share behaviours. This increases the amount of variance that the model can account for from \textasciitilde 6\% to \textasciitilde 29\% (McFadden's R\textsuperscript{2} = .286).}
		\label{table:results:logregtwo} 	
		
		\centering
		\footnotesize

		\begin{tabularx}{\linewidth}{Xccl} 
			
			\toprule
			\textbf{Coefficient} & \textbf{Value} & \textbf{SE} & \hspace{.4cm}\textbf{p} 
			\hspace{.6cm}
			\\ 
			
			\midrule
			
		\end{tabularx}

		\vspace{.1cm}
		
		\flushleft{\textbf{Step 3:} Contextual, Demographic \& Share Proportions}
		
		\begin{tabularx}{\linewidth}{Xccl}

			\hspace{.5cm}Intercept			        
			& -3.30 & 0.18 & 0.000 *** \\ 
			
			\hspace{.5cm}Data type & ~ & ~ & ~ \\
			\hspace{1cm}Checkin					   
			& 0.57 & 0.16 & 0.000 *** \\ 
			\hspace{1cm}Like 						   
			& 1.41 & 0.15 & 0.000 *** \\  
			\hspace{1cm}Note 						 
			& 0.80 & 0.15 & 0.000 *** \\ 
			\hspace{1cm}Photo 						
			& 0.37 & 0.14 & 0.010 * \\ 
			
			\hspace{.5cm}Education & ~ & ~ & ~ \\
			\hspace{1cm}High School  					   
			& 0.05 & 0.14 & 0.719  \\ 
			\hspace{1cm}Undergraduate Degree 						
			& -0.09 & 0.12 & 0.477  \\ 
			\hspace{1cm}Postgraduate Degree 			    
			& 0.47 & 0.14 & 0.001 ** \\ 
			
			\hspace{.5cm}Number of Friends
			& 0.00 & 0.00 & 0.007 ** \\ 
			
			\hspace{.5cm}Total Share Proportion
			& 2.19 & 0.26 & 0.000 *** \\ 
			
			\hspace{.5cm}Audience Share Proportion
			& 2.63 & 0.20 & 0.000 *** \\

		\end{tabularx}
		
		\vspace{.1cm}
		\hfill McFadden's R\(^2\) = .286
		\vspace{.1cm}

		\hrulefill	
		
		\vspace{.1cm}
		
		\centering
		* p \textless~.05;\hspace{1.2em} 
		** p \textless~.01;\hspace{1.2em} 
		*** p  \textless~.001

	\end{table}
}

%
%
%
%

\newcommand{\tblmodellist}{
\renewcommand{\arraystretch}{1.3}
\begin{table*}[!b]
	\caption{Models evaluated on the test set. Many of the models perform comparatively when using the ROC curve (AUC) or the F\textsubscript{1} score as a performance metric, however, the na\"{\i}ve Bayes classifier has the highest specificity -- suggesting fewer data leaks.}
	\label{table:results:modellist} 
	
	\centering
	\footnotesize
	\begin{tabularx}{\textwidth}{Xlllllll} 
	
		\toprule
		\textbf{Model} & \textbf{Accuracy} & \textbf{Precision} & 
		\textbf{Sensitivity} & 
		\textbf{Specificity} & \textbf{F\textsubscript{1} Score} & \textbf{AUC} \\ 
		
		\midrule
		
		Na\"{\i}ve Bayes Classifier 
		& 0.765 & 0.827 & 0.737 & 0.802 & 0.779 & 0.849 \\ 
		
		Support Vector Machine (RBF) 
		& 0.773 & 0.807 & 0.786 & 0.758 & 0.796 & 0.840 \\ 
		
		Logistic Regression 
		& 0.755 & 0.776 & 0.794 & 0.705 & 0.785 & 0.830 \\ 
		
		Neural Network (MLP) 
		& 0.770 & 0.804 & 0.782 & 0.755 & 0.793 & 0.827 \\ 
		
		Random Forest 
		& 0.730 & 0.786 & 0.715 & 0.749 & 0.749 & 0.803 \\ 
		
		k-Nearest Neighbour 
		& 0.690 & 0.708 & 0.764 & 0.595 & 0.735 & 0.732 \\

		 \bottomrule
	\end{tabularx}
\end{table*}
}

%
%
%
%
%

\newcommand{\tbloptimisationlist}{
	\renewcommand{\arraystretch}{1.3}
	\begin{table*}[!b]
		\caption{Na\"{\i}ve Bayes models evaluated on the test set. Models either use the full or minimised formulas, either with or without threshold adjustments. The threshold-adjusted models have slightly lower accuracy for the full and minimised models. See the Discussion section for recognised limitations of the Minimised + Threshold Ajusted model.}
		\label{table:results:optimisationlist} 
		\centering
		\footnotesize
		\begin{tabularx}{\textwidth}{XXlllllll} 
			
			\toprule
			\textbf{Formula} & \textbf{Cost Adjustment} & \textbf{Accuracy} & 
			\textbf{Precision} & \textbf{Sensitivity} & \textbf{Specificity}
			& \textbf{F\textsubscript{1} Score} & \textbf{AUC} \\ 
			
			\midrule
			
			Full & None
			& 0.765 & 0.827 & 0.737 & 0.802 & 0.779 & 0.849 \\ 
			
			Full & Threshold
			& 0.698 & 0.894 & 0.525 & 0.920 & 0.661 & 0.849 \\ 
			
			Minimised & None
			& 0.745 & 0.737 & 0.848 & 0.612 & 0.789 & 0.817 \\ 
			
			Minimised & Threshold
			& 0.728 & 0.914 & 0.570 & 0.931 & 0.702 & 0.817 \\

			\bottomrule
		\end{tabularx}
	\end{table*}
}

%
%
%
%

\newcommand\MyBox[1]{
	\fbox{\lower0.75cm
		\vbox to 1.7cm{\vfil
			\hbox to 1.7cm{\hfil\parbox{1.4cm}{\centering#1}\hfil}
			\vfil}
	}
}

\newcommand{\tblconfusionmatrix}{
	\begin{table}[!t]
	\caption{Confusion matrix for the threshold adjusted na\"{\i}ve Bayes model using the full set of predictors, evaluated using the validation set of participants. 
	}
	\label{table:results:confusionmatrix} 
	
	\small
	\noindent
	\renewcommand\arraystretch{1.3}
	\setlength\tabcolsep{0pt}
	\begin{tabular}{c >{\bfseries}r @{\hspace{0.7em}}c @{\hspace{0.4em}}c 
	@{\hspace{0.7em}}l}
		\multirow{10}{*}{\rotatebox{90}{\parbox{1.1cm}{\bfseries\centering Actual\\ 
		value}}} & 
		& \multicolumn{2}{c}{\bfseries Predicted value} & \\
		& & \bfseries Share & \bfseries Do not share & \bfseries total \\
		& Share & \MyBox{129} & \MyBox{265} & 394 \\[2.4em]
		& Do not share & \MyBox{12} & \MyBox{393} & 405 \\
		& total & 141 & 658 & ~

	\end{tabular}
	\end{table}
}